\documentclass[letter,english]{article}

\usepackage{graphics}
\usepackage{epstopdf}
\usepackage{overcite}
\usepackage{graphicx}
\usepackage{amsmath}
\usepackage{amssymb}
\usepackage{bm}
\usepackage{epsfig}
\usepackage{subfigure}
\usepackage{graphpap}

\begin{document}
% **********************************************************
% * TITLE, AUTHORS AND AFFILIATIONS                        *
% **********************************************************
\title{Optical properties of emeraldine salt polymers from ab initio calculations:
comparison with recent experimental data}
\author{Renato Colle$^1$\footnote{colle@sns.it}\and
Pietro Parruccini$^1$\footnote{parrucci@sns.it}\and
Andrea Benassi$^{2,3}$\footnote{benassi.andrea@unimore.it}\and
Carlo Cavazzoni$^{2,4}$\footnote{cavazzoni@cineca.it}}
\date{}
\maketitle
\begin{description}
\item[$^1$] Dipartimento di Chimica Applicata e Scienza dei Materiali, Universit\`a di
Bologna, via Saragozza,8 I-40136 Bologna, Italy
\item[$^2$] INFM-National Research Center on nanoStructures and bioSystems at Surfaces (S3)
\item[$^3$] Dipartimento di Fisica, Universit\`a di Modena e Reggio Emilia,
Via Campi 213/A, I-41100 Modena, Italy
\item[$^4$] High Performance Computing Department CINECA, I-40136 Bologna, Italy
\end{description}
% **********************************************************
% * ABSTRACT                                               *
% **********************************************************
\begin{abstract}
We present absorption coefficient $\alpha(\omega)$, transverse
dielectric function $\epsilon(\omega)$, optical conductivity
$\sigma(\omega)$, and reflectance $R(\omega)$ calculated for an
emeraldine salt conducting polymer in its crystalline three-dimensional
polaronic structure. We utilize Kohn-Sham DFT electronic
wavefunctions and energies implemented in the expression of the macroscopic
transverse dielectric function {\bf in the framework of the band
theory without the electron-hole interaction.}
Contributions of intra-band
transitions are taken into account by adding a Drude-like term to
the dielectric function calculated ab initio. Comparison with
optical properties, recently measured on high-quality emeraldine
salts [Nature, {\bf2006}, 441, 65], and with optical absorption
spectra, recorded on other emeraldine salts, is very satisfactory.
The calculated spectra are discussed in terms of energy-band
structure, density of states, inter- and intra-band transitions and
transverse dielectric function.
\end{abstract}
\textbf{pacs: }{ 71.20.Rv, 72.80.Le, 78.40.Me, 78.66.Qn}
\section{Introduction}
Conducting polymers are a new~\cite{SLMCH,CFPHSL,RA} and exciting
field of research at the boundary between chemistry and condensed
matter physics. They address a number of theoretical questions of
enormous interest~\cite{hee01,FG} and open the possibility of
several important technological applications\cite{REW,SCS,PMF,REV,WTC,lwtJACS99,pyzSC95,srMCP02}.

Conducting polymers are conjugated systems in which the electronic
delocalization, due to $\pi-$bonding, provides conditions for
charge mobility along the backbone of the polymer chains.
Furthermore, the electron attraction with the nuclei of the
neighboring chains makes also possible interchain electron
transfer, favored by the crystalline order of the structure. The
conducting properties of these polymers depend on the
characteristics of their repeat unit and can be reversibly
modified from insulator to metal by introducing p-type or n-type
carriers via electrochemical doping~\cite{nmhJCC79,pyzSC95} or
acid-base chemistry (protonation)\cite{slhSM86,cmSM86,mreSM87,egzSM87}.

Among conducting polymers, polyanilines (PANI)~\cite{MDE90283}
have given the first example of a conjugated polymer converted by
the solvent (a protonic acid) from a semiconductor form
(emeraldine base EB) to a conducting form (emeraldine salt ES),
which can be processed directly into the metallic salt from the
concentrated acid solution\cite{cshSM89}. The solubility of the
resulting PANI complex in common organic solvents\cite{cshSM92}
has opened the way to processing the conducting polymer with other
commercial polymers into films and polymer blends\cite{ycsSM92}.
The existence of phase segregation into metallic and non-metallic
phases\cite{egzSM87,grmSC87} and of crystalline and amorphous
regions in the ES powders and films has been evidenced by various
experimental studies\cite{grmSC87,jljPRB89,pjeMA91,
lpsMA92,wmMA98,cfgMA99,mfcMA99}, together with the existence of
two classes (ES-I and ES-II) of the conducting
polymer\cite{jljPRB89,pjeMA91,lpsMA92} obtained from acidification
of EB by HCl.

In a single chain representation of the polymer structure, the
repeat unit of EB is $[(1A)(2A)]_n$, with $(1A)\equiv(B-NH-B-NH-)$
and $(2A)\equiv(B-N=Q=N-)$, being B and Q a benzenoid and a
quinoid ring, respectively. Protonation of this
unit produces a structural change of the chain, without changing
however the number of electrons\cite{MDE9053,walJACS87}, and
leads to emeraldine salts with chain repeat unit:
$[(1S)^{\cdot+}(C)^-]_n$, where
$(1S)^{\cdot+}\equiv(B-NH-B-NH-)^{\cdot+}$ and $(C)^-$ is the
counterion (e.g. $Cl^-,ClO_4^-,HSO_4^-,CS^-,...)$. The repeat unit
$[(1S)^{\cdot+}]_n$ has a unpaired electron in a half-filled band,
that leads to a metallic state described as a polaronic
metal\cite{rgwPRL88}.

The characterization of the emeraldine salts as conducting
polymers has been extensively performed by probing their
electronic structure through optical studies (mainly absorption
spectra and photoinduced absorption spectra) of films cast from
acid solutions\cite{egzSM87,cshSM89,rgwPRL88} and of the polymer
dissolved in acid solutions\cite{cshSM89,frgSM00}. The
experimental spectra have been interpreted utilizing VEH
band-structure calculations\cite{bcwJCP86,sbePRL87} performed on a
single-chain model of the polymer, without calculating, however,
dielectric function and related optical properties. The VEH
technique has been also used to characterize chain geometry and
electronic modifications induced in isolated PANI chains by doping\cite{lbePRB95}.
More recently, DFT calculations have been
performed\cite{assJACS05} to compare structural properties and
relative energies of isolated EB and ES polymer chains involved in
the mechanism of HCl/EB protonation. In a recent
paper\cite{ccfgPRB06}, the HCl protonation process of the EB-II
polymer to give the ES-II crystalline salt has been simulated
using the Car-Parrinello Molecular Dynamics\cite{cpPRL85,mhREV00}.
We have shown that a $Pc2a$ lattice structure of polymer chains,
drastically rearranged with respect to those of
EB-II\cite{ccfgPRB04} and with chlorine ions distributed in
polaronic arrangement, is in good agreement with the available
X-ray experimental data\cite{jljPRB89,pjeMA91} recorded on ES-II
polymers.

Optical, magnetic and transport properties of polyanilines in
their metallic state have been intensively studied in the last
decade\cite{wlsPRL91,WSME92,RCMH93,LHC93,jlpPRB98,TZHM02,hee02,LEE,LH02}.
The measured low-frequency dependence of the optical conductivity
$\sigma(\omega)$ and of the real part of the dielectric function
$\epsilon_1(\omega)$, and the low temperature behavior of the
resistivity have been found not those expected for conventional
metals. Few months ago, truly metallic polymers characterized by
classic metallic transport data have been eventually
obtained~\cite{nature06} from samples of PANI-EB prepared with the
self-stabilized dispersion polymerization (SSDP)
technique\cite{llll05}, and doped with camphor sulphonic acid
(CSA) to give PANI-CSA salts. Measures of resistivity, down to
T$\sim5K$, and of $\sigma(\omega),~\epsilon_1(\omega)$ and
reflectance $R(\omega)$ below $2000~cm^{-1}$ have confirmed the
true metallic character of protonated PANI when prepared from
high-quality unprotonated PANI-EB samples with low density of
structural defects. This decisive step in the characterization of
ES polymers as true metallic systems shows the importance of an
ordered structure of regular chains. This fact makes also the
theoreticians more confident in modeling the crystalline regions
of such systems as three-dimensional perfect crystals of mutually
interacting\cite{LH02} regular chains, a model that is
particularly suitable for ab initio calculations of the electronic
and optical properties which characterize the \lq\lq metallic
state" of these polymers.

In this paper, we calculate absorption coefficient
$\alpha(\omega)$, complex dielectric function $\epsilon(\omega)$,
optical conductivity $\sigma(\omega)$, and reflectance $R(\omega)$
of a three-di\-men\-sional crystal of ES-II polymer in its
polaronic structure\cite{ccfgPRB06}. We utilize Kohn-Sham DFT electronic
wavefunctions and energies implemented in the expression of the macroscopic
transverse dielectric function in the framework of the band
theory without the electron-hole interaction\cite{rlPRB00}. In this calculation, we have included only vertical band-to-band transitions. To take into account also contributions of the intra-band transitions, we have added, to the dielectric function calculated ab initio,
a Drude-like term appropriate for metallic systems. The
resulting global expression of $\epsilon(\omega)$ allows us to
reproduce quite satisfactorily the measured $\omega-$dependence of
the main optical properties in the range of frequencies from
$\sim0.2~$eV to above 4~eV.

{\bf We point out that several first-principles calculations of electronic and optical properties of crystalline conjugated polymers\cite{rcbmPRL02} and oligo\-mers\cite{haPRB05,ahspCP06} have been performed using also correlated many-body techniques. To our knowledge, however, this is the first attempt to construct, from ab initio calculations, optical spectra of a crystalline conducting polymer obtained by chemical doping of a conjugated polymer.}

In Sec. \ref{method}, we present the quantum-mechanical
expressions used for calculating the optical quantities of
interest, and we describe the computational procedures used for
their evaluation. In Sec. \ref{resa}, we present the optical
absorption spectrum calculated for a 3D crystal of
ES-II\cite{ccfgPRB06} in its polaronic structure obtained from HCl
protonation of EB-II\cite{ccfgPRB04}, and we discuss the absorption
spectrum in terms of band-structure and density of states
resulting from KS-DFT calculations. We also compare the optical
absorption spectrum calculated for ES-II with that calculated for
the EB-II polymer\cite{ccfgPRB04}, and with experimental spectra
recorded on solid samples of the two polymers\cite{cshSM89,rgwPRL88,egzSM87}.
Finally, in Sec. \ref{resb}, we calculate the optical spectra of the real part of the dielectric function, optical conductivity and reflectance of ES-II, identifying the spectral contributions due to the inter- and intra-band electronic transitions. The calculated spectra are compared with those recently measured
on PANI-CSA salts~\cite{nature06}. Conclusions are drawn in Sec.\ref{conc}.

\section{Method}
\label{method} We have carried out KS-DFT calculations of
energy-band structure, density of states and electronic
wavefunctions of a perfect 3D crystal of ES-II in its polaronic
structure with lattice parameters and chains geometry taken from
Ref.\cite{ccfgPRB06}. To this end, we have used the package of programs
$Quantum-Espresso$\cite{esp}, which performs ab initio calculations of
ground-state energy, energy gradient, electronic wave-functions
and properties of periodic systems. These codes use plane-wave
expansion for the single particle wavefunctions, pseudopotentials
(ECP) for the core electrons, and various energy functionals for
the exchange-correlation potential. In our KS-DFT calculations we
have used a kinetic energy cut-off of 70 Rydberg, different grids of $k-$points
up to a maximum of 2200 points, the
norm-conserving ECP of Martin and Troullier~\cite{tmPRB91}, and
the Becke-Lee-Yang-Parr (BLYP) exchange-correlation
functional~\cite{bPRA88, lypPRB88}.
Energy bands and wavefunctions obtained from these calculations
have been used for evaluating the macroscopic transverse dielectric function, whose real and imaginary parts enter into the definition of the main optical properties of a condensed system.

For an incident light of frequency $\omega$, wave vector ${\bf q}$
and polarization direction $\hat e$, {\bf the macroscopic transverse dielectric
function, defined in the framework of the band
theory without the electron-hole interaction,} is given by
\begin{equation}\label{E1}
\epsilon({\bf q},\hat e;\omega)=1+\frac{4\pi\hbar^2e^2}{
m^2V}\sum_{c,{\bf k_c}}\sum_{v,{\bf k_v}}\frac{(n_{v {\bf
k_v}}-n_{c{\bf k_c}})}{(E_{c {\bf k_c}}-E_{v{\bf k_v}})^2}
\frac{|<\Psi_{c {\bf k_c}}|e^{i{\bf q}\cdot {\bf r}}\hat e\cdot {\bf
p}|\Psi_{v {\bf k_v}}>|^2}{(E_{c {\bf k_c}}-E_{v {\bf k_v}}-\hbar\omega-i\eta)}
\end{equation}
with $\eta\rightarrow 0^+$. In Eq.(\ref{E1}), $\{\Psi_{j {\bf k_j}}({\bf r})\}$
are crystalline orbitals of energy $E_{j {\bf k_j}}$ and of occupation
number $n_{j {\bf k_j}}$, and $V$ is the crystalline volume given by
the product of the cell volume times the number of cells.

In dipole approximation $(q=0)$, and expanding the crystalline
orbitals in terms of plane waves of the reciprocal lattice
for each ${\bf k}$ of the first Brillouin zone: $\Psi_{j {\bf
k}}({\bf r})=\frac{1}{\sqrt V}\sum_{{\bf g}\in RL}C_j({\bf k}+{\bf g})e^{i({\bf k}+{\bf
g})\cdot{\bf r}}$, the transition matrix element in Eq.(\ref{E1})
becomes
\begin{align}\label{E2}
<\Psi_{c{\bf k_c}}|\hat e\cdot{\bf p}|\Psi_{v{\bf
k_v}}>&=\hbar\delta_{{\bf k_c},{\bf k_v}}\sum_{{\bf g}\in RL}\hat
e\cdot({\bf k_v}+{\bf g})C^*_c({\bf k_c}+{\bf
g})C_v({\bf k_v}+{\bf g})\nonumber=\\
&=M_{c,v}(\hat e,{\bf k})\qquad\qquad
{\bf k}={\bf k_c}={\bf k_v}
\end{align}
Note that, in our calculations, the core electrons are not
included explicitly, but only through ECP, thus the crystalline
orbitals, obtained from the solution of the KS-equations, are
\lq\lq pseudo" wavefunctions, which overlap the \lq\lq exact" ones
only beyond the \lq\lq core radius" of the various atoms. The
corrections to the dipole matrix elements due to the ECP non-local
terms in the hamiltonian have been neglected in our calculations,
together with local field effects due to the inhomogeneity of the crystal\cite{baroniresta}.

Taking the limit of small, but non vanishing $\eta$ in Eq.(\ref{E1}), a procedure that
accounts for the non-monochromaticity of the incident light and the finite lifetime of
the excited levels, we obtain the following Drude-Lorentz expression for the
dielectric function:
\begin{equation}\label{E3}
\epsilon(\hat e;\omega)=1+\omega_p^2~\sum_{c,v}\sum_{{\bf k}}~\frac{f_{v{\bf k}}^{c{\bf k}}
(\hat e)}{[(\omega_{c{\bf k}}-\omega_{v{\bf k}})^2-\omega^2]-i~\Gamma\omega}
\end{equation}
where $\omega_{c{\bf k}}=E_{c{\bf k}}/\hbar$ and $\Gamma=2\eta/\hbar$. Furthermore,
$\omega_p=\sqrt{\frac{4\pi e^2}{m}N}$ is the free-electron
plasma frequency with $N$ the number of electrons per unit volume, and
$f^{c{\bf k}}_{v{\bf k}}(\hat e)$ is the oscillator
strength of the $|\Psi_{v{\bf k_v}}>\rightarrow|\Psi_{c{\bf k_c}}>$
transition:
\begin{equation}\label{E4}
f_{v{\bf k}}^{c{\bf k}}(\hat e)=\frac{2~n_{v{\bf k}}}{\hbar m NV}~\frac{|M_{c,v}(\hat e,{\bf k})|^2}
{\omega_{c{\bf k}}-\omega_{v{\bf k}}}
\end{equation}

The real and imaginary parts of $\epsilon(\hat e,\omega)$ are,
respectively,
\begin{equation}\label{E5}
\epsilon_1(\hat e;\omega)~=~1+\omega_p^2 \sum_{c,v}\sum_{{\bf k}}~
f^{c{\bf k}}_{v{\bf k}}(\hat e)~\frac{[(\omega_{c{\bf k}}-\omega_{v{\bf k}})^2-\omega^2]}
{[(\omega_{c{\bf k}}-\omega_{v{\bf k}})^2-\omega^2]^2+(\Gamma\omega)^2}
\end{equation}
and
\begin{equation}\label{E6}
\epsilon_2(\hat e;\omega)~=~\omega_p^2 \sum_{c,v}\sum_{{\bf k}}
~f^{c{\bf k}}_{v{\bf k}}(\hat e)~\frac{\Gamma\omega}
{[(\omega_{c{\bf k}}-\omega_{v{\bf k}})^2-\omega^2]^2+(\Gamma\omega)^2}
\end{equation}

For computational reasons, we have included only vertical
band-to-band transitions in the implementation of
Eqs.(\ref{E5}-\ref{E6}), thus disregarding the intra-band
electronic transitions that give an important contribution to the
low frequency spectral region of the metallic systems. A
simplified way of taking into account this contribution is through
the Drude theory of metals, which suggests to complement the real and
imaginary parts of the dielectric function, calculated including only
inter-band transitions, as follows
\begin{align}\label{E7}
\overline{\epsilon_1}(\hat e;\omega)&=\epsilon_1(\hat
e;\omega)-\omega_p^2 \frac{f_D}{\omega^2+\gamma^2}\nonumber\\
\overline{\epsilon_2}(\hat e;\omega)&=\epsilon_2(\hat
e;\omega)+\omega_p^2\frac{f_D~\gamma}{\omega(\omega^2+\gamma^2)}
\end{align}
where $f_D$ and $\gamma$ represent phenomenological parameters for
the free carriers of the metallic system. In fact, we have used
these parameters to best fit the frequency dependence of the real
part of the dielectric constant measured on PANI-CSA polymers
\cite{nature06}.

We point out that this way of including the effects of the
intra-band transitions represents an a posteriori,
phenomenological correction of the ab initio dielectric function.
Therefore, it does not assure the rigorous fulfilment either of
the dispersion relations or of the sum rule. In the case of the
ES-II polymer, we have found that our ab initio calculations,
which take into account only inter-band transitions, give
$\Sigma_{c,v}\Sigma_{{\bf k}} f^{c {\bf k}}_{v{\bf k}}=0.885$, while
the best fit value obtained for the strength of the Drude
oscillator is $f_D=0.014$. Adding these oscillator strengths,
we obtain a 90\% fulfillment of the corresponding sum rule.
Furthermore, our best fit value for the Drude damping parameter is
$\hbar\gamma=0.55$ eV. This value, if used to estimate the d.c.
conductivity of the polymer, gives
$\sigma(\omega=0)=\omega_p^2~\frac{f_D~\gamma}{4\pi}\simeq1187~S~cm^{-1}$,
in good agreement with the experimental value, measured at T=300
K on PANI-CSA polymers\cite{nature06}, that is $\sigma(\omega=0)\simeq1100~S~cm^{-1}$.

The optical properties calculated in this paper are the following:
\begin{itemize}
\item {\it real $(+)$ and imaginary $(-)$ parts of the complex
refractive index} :
\begin{equation}\label{E9}
n_{\pm}(\hat
e;\omega)~=\sqrt{\frac{1}{2}\sqrt{\epsilon_1(\hat
e;\omega)^2+\epsilon_2(\hat
e;\omega)^2}\pm\frac{1}{2}\epsilon_1(\hat e;\omega)}
\end{equation}
\item {\it optical absorption spectrum} :
\begin{equation}\label{E10}
\alpha(\hat e;\omega)~=\frac{\omega}{c~n_+(\hat
e;\omega)}~\epsilon_2(\hat e;\omega)=\frac{2\omega}{
c}n_-(\hat e;\omega)
\end{equation}
\item {\it optical conductivity} :
\begin{equation}\label{E11}
\sigma(\hat e;\omega)=\frac{\omega}{
4\pi}~\epsilon_2(\hat e;\omega)
\end{equation}
\item {\it reflectance} :
\begin{equation}\label{E12}
R(\hat e;\omega)~=\frac{[n_+(\hat e;\omega)-1]^2+n_-(\hat e;\omega)^2
}{[n_+(\hat e;\omega)+1]^2+n_-(\hat e;\omega)^2}
\end{equation}
\end{itemize}

%%%%%%%%%%%%%%%%%%%%%%%%%%%%%%section III%%%%%%%%%%%%%%%%%%%%%%%%%%%%%%%
\section{Results}

For the ES-II polymer, we have used a $Pc2a$ orthorhombic cell
(lattice parameter: $a=7.1$~\AA, $b=7.9$~\AA, $c=20.84$~\AA)
with two strands of polymer chains inside, each one with four
$(-C_6H_4NH-)$ groups and two adjacent chlorine atoms in polaronic
positions (see Ref.\cite{ccfgPRB06}). The structural and electronic results
obtained for ES-II\cite{ccfgPRB06} show that each polymer
chain consists of a repeat unit $[(1S)^{\cdot+}(Cl)^-]_n$, and the
two strands of polymer chains in the cell are mutually shifted by
one complete ring along the chain backbone. Thus, we have
consistently reduced the $c$-axis of our simulation cell by a
factor two ($c=10.42$~\AA), including two strands
of polymer chains with repeat unit $[(1S)^{\cdot+}(Cl)^-]_n$ and
geometrical parameters taken from Table I and II of
Ref.\cite{ccfgPRB06}.

For the EB-II polymer, we have used the  $Pbcn$ orthorhombic cell
defined in Ref.\cite{ccfgPRB04} (lattice parameter: $a=7.80$~\AA,
$b=5.75$~\AA, $c=20.25$~\AA) with two strands of polymer chains inside,
each one with three phenilene rings, one quinoid ring, two amine and two
imine nitrogen atoms, and with geometrical parameters taken from Table I
and II of Ref.\cite{ccfgPRB04}.

\subsection{Optical absorption spectra}\label{resa}

\begin{figure}
\centering
\includegraphics[width=\linewidth]{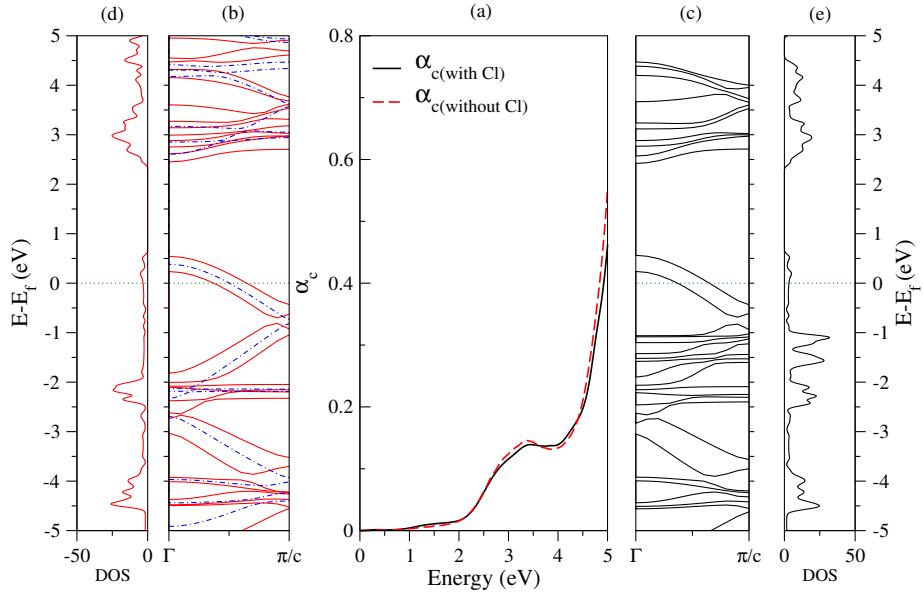}
\caption{\label{fig1} {\bf(a)} Absorption coefficient,
$\alpha(\hat c,E)$, for a linearly polarized light along the
direction of the polymer chains. The dashed line corresponds to
the absorption spectrum of the system without counterions. ({\bf
b, c}) Energy-band structures, measured from the Fermi energy and
plotted along the $c-$axis of the reciprocal space, of the ES-II
polymer without counterions ({\bf b}) and with counterions in
polaronic positions ({\bf c}); the dashed-dot lines in ({\bf b})
are the energy-bands of an isolated polymer chain of ES-II
without counterions. ({\bf d, e}) Total densities of states (DOS)
corresponding to the energy-band structures shown in ({\bf b,
c}), respectively.}
\end{figure}

In Fig.\ref{fig1}{\bf a}, we plot the absorption coefficient
$\alpha(\hat c;\omega)$ calculated with the ab initio components
$\epsilon_{1,2}(\hat c;\omega)$ of the dielectric function. In
Eqs.(\ref{E5}-\ref{E6}), we have used $\hbar\Gamma=0.3$ eV and a
linearly polarized light parallel to the $c-$axis of the cell,
i.e. to the direction of the chains backbone. Note that
$\alpha(\hat c,\omega)$ is the largest component of the absorption
coefficient  $\alpha(\hat e,\omega)$; thus, even plotting the
average value of the three components of  $\alpha(\hat e,\omega)$,
the resulting spectrum mantains almost the same profile.
In Fig.\ref{fig1}{\bf a}, we have also plotted the absorption
spectrum calculated for the ES-II polymer without counterions, i.e.
with two strands of chains with repeat unit $[(1S)^{\cdot+}]_n$ inside the same
crystalline cell. Comparing the two plots of  Fig.\ref{fig1}{\bf
a}, we see that the spectral contribution of the counterions is
negligible, in agreement with the fact that absorption spectra
recorded on samples with different counterions have almost
identical profiles.

In Figs.\ref{fig1}{\bf b} and \ref{fig1}{\bf c}, we plot the
energy-band structure of the system without and with counterions,
respectively, and, in Figs.\ref{fig1}{\bf d} and \ref{fig1}{\bf
e}, we plot the corresponding densities of states (DOS). Comparing these
figures, we see that the only contributions of the  $Cl^-$
counterions are the flat bands due to their $3p$ levels in the
region between 1-2 eV below the Fermi level. (Note that the Fermi
energy has been calculated by integrating the DOS to the proper
number of electrons.) {\bf The energy-band structure with the characteristic doubling of the bands
makes evident the presence of two interacting chains in the unit cell.} For comparison, in
Fig.\ref{fig1}{\bf b} we have plotted also the band structure of an
isolated chain of the polymer without counterions. We
observe that, considering only vertical transitions,
Fig.\ref{fig1}{\bf b} predicts the onsets of an $intra-chain$
absorption band around 1 eV and of an $inter-chain$ absorption band
around 1.5 eV. Looking, however, to the absorption spectrum of
Fig.\ref{fig1}{\bf a}, obtained with the explicit calculation of
the transition matrix elements in $\epsilon(\hat c,\omega)$, we
see that the intensity of the transitions below 2 eV is quite low;
the absorption spectrum, indeed, has its first relevant peak
around 3.5 eV, where both $intra-$ and $inter-chain$
transitions from partially and fully occupied bands give their
contributions.

%
%      Figura 1
%

\begin{figure}
\centering
\includegraphics[width=\linewidth]{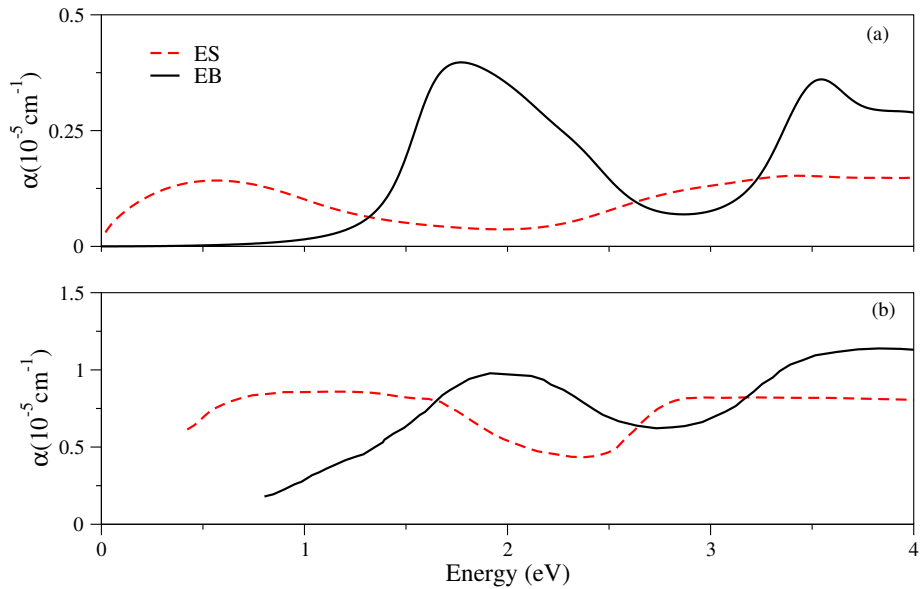}
\caption{\label{fig2} Comparison between calculated {\bf(a)} and
experimental ({\bf b}) absorption spectra of ES and EB  polymers.
Figure \ref{fig2}(b) has been adapted from Fig. 3 of
Ref.\cite{cshSM89}.}
\end{figure}

In Fig.~\ref{fig2}, we compare an experimental absorption
spectrum\cite{cshSM89} of ES, recorded on polyaniline film
spin-cast from sulphuric acid solution and subsequently
equilibrated with aqueous HCl solution ($pH=0.15$), with the \lq\lq
theoretical" absorption spectrum calculated using the dielectric
function ${\overline \epsilon}(\hat c,\omega)$ that includes the
Drude term for intra-band transitions. We see that the \lq\lq
theoretical" spectrum shows, now, a broad absorption region below
1.5 eV, slightly red-shifted with respect to the corresponding
region in the experimental spectra (see also Fig.1b of
Ref.\cite{rgwPRL88} and Fig.4a of Ref.\cite{egzSM87}). Nevertheless,
the global profile of the \lq\lq theoretical" spectrum follows quite satisfactorily the
experimental one.

In Fig.\ref{fig2}, we also compare the absorption spectrum
measured on film of EB polymer\cite{cshSM89} with the \lq\lq
theoretical" absorption spectrum calculated for the EB-II polymer
of Ref.\cite{ccfgPRB04}. Since EB is a semiconductor (direct gap
$\sim1.4$ eV), we have used the dielectric function $\epsilon(\hat
c,\omega)$ without Drude correction, and {\bf we have applied a scissor
operator of 0.8 eV, i.e. a k-independent self-energy correction that shifts rigidly the
conduction bands to enlarge the band gap underestimated
in the DFT calculations\cite{gssPRB88}}. We see that the \lq\lq
theoretical" spectrum of EB reproduces quite satisfactorily the
two characteristic peaks around 2 eV and 3.5 eV. The first one is
ascribed to a charge-transfer transition associated with
excitation from benzenoid to quinoid rings, and the latter to a
$\pi \rightarrow\pi^*$ band-gap absorption\cite{rgwPRL88,mglPRB90}.
Upon protonation of EB to give ES, the first peak vanishes and two absorption regions
appear centered around 1 eV and 3 eV, respectively.

We observe that the spectral profiles of the absorption spectra calculated for
EB-II and ES-II are in substantial agreement with the experimental ones,
even if the ratio of the ES/EB intensities is slightly underestimated in the theoretical spectrum.

%
%      Figura 2
%

\subsection{Dielectric function, optical conductivity and reflectance}\label{resb}

In this section, we present the $\epsilon_1(\hat c,\omega)$,
$\sigma(\hat c,\omega)$ and $R(\hat c,\omega)$ spectra calculated
for the ES-II crystalline polymer of Ref.\cite{ccfgPRB06}, and we
compare these spectra with those recorded on PANI-CSA
salts\cite{nature06} prepared from SSDP samples of PANI-EB. The
comparison is based on the assumption that lattice structure and polymer chains
of PANI-CSA are not too different from those of ES-II\cite{ccfgPRB06}, and the specific contribution of the counterions to the optical spectra of these polymers can be
neglected. We point out also that our comparison with the experimental spectra
does not include the $\omega-$region below 0.2 eV, where phonon
contributions are dominant.

\begin{figure}[ht]
\centering
\includegraphics[width=0.8\linewidth]{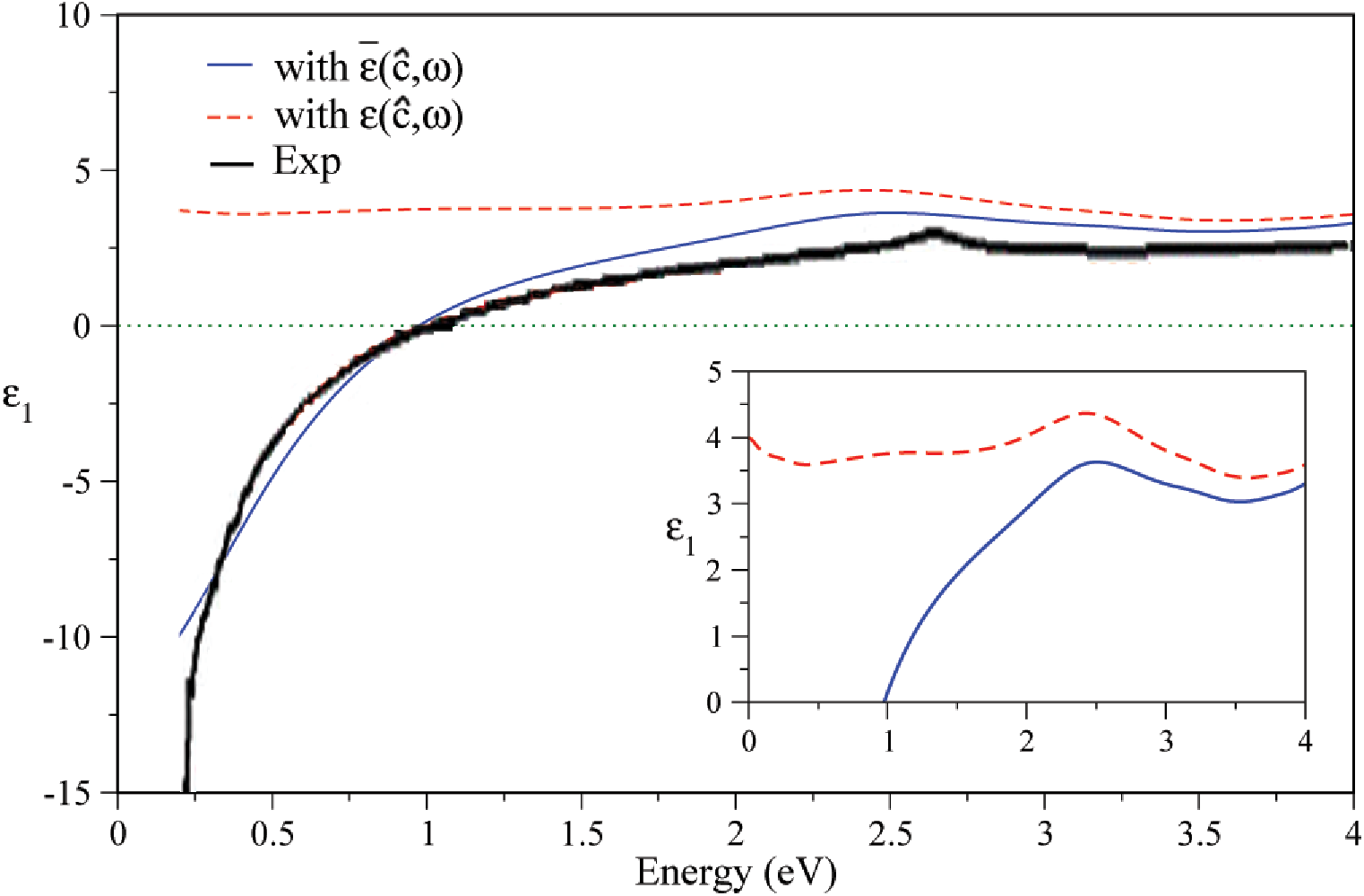}
\caption{\label{fig3}  Measured spectrum of the real part of the
dielectric function of PANI-CSA salts\cite{nature06}, compared with the spectra calculated with and without the Drude term. The inset shows details of the calculated curves in
the frequency region of interest.}
\end{figure}

In Fig.~\ref{fig3}, we compare the measured and calculated spectra of the
real part of the dielectric function;
we have plotted separately $\epsilon_1(\hat c;\omega)$ and ${\overline
\epsilon_1}(\hat c;\omega)$ to make evident the contributions of
the inter- and intra-band transitions. An appropriate choice of
the $f_D$ and $\gamma$ parameters in Eq.(\ref{E7})
makes ${\overline \epsilon_1}(\hat c;\omega)$ to cross zero around
$\hbar\omega\sim1$ eV, as in the experiment, and to remain negative
down to $\hbar\omega\sim 0.2$ eV (the limit value for our comparison
with the experiment). We see that the characteristic bump around 2.7 eV
in the experimental spectrum is slightly red-shifted by the ab initio component
$\epsilon_1(\hat c;\omega)$ of the dielectric function, but the overall
agreement between calculated and measured spectra is remarkable.

%
%      Figura 3
%

\begin{figure}[ht]
\centering
\includegraphics[width=0.8\linewidth]{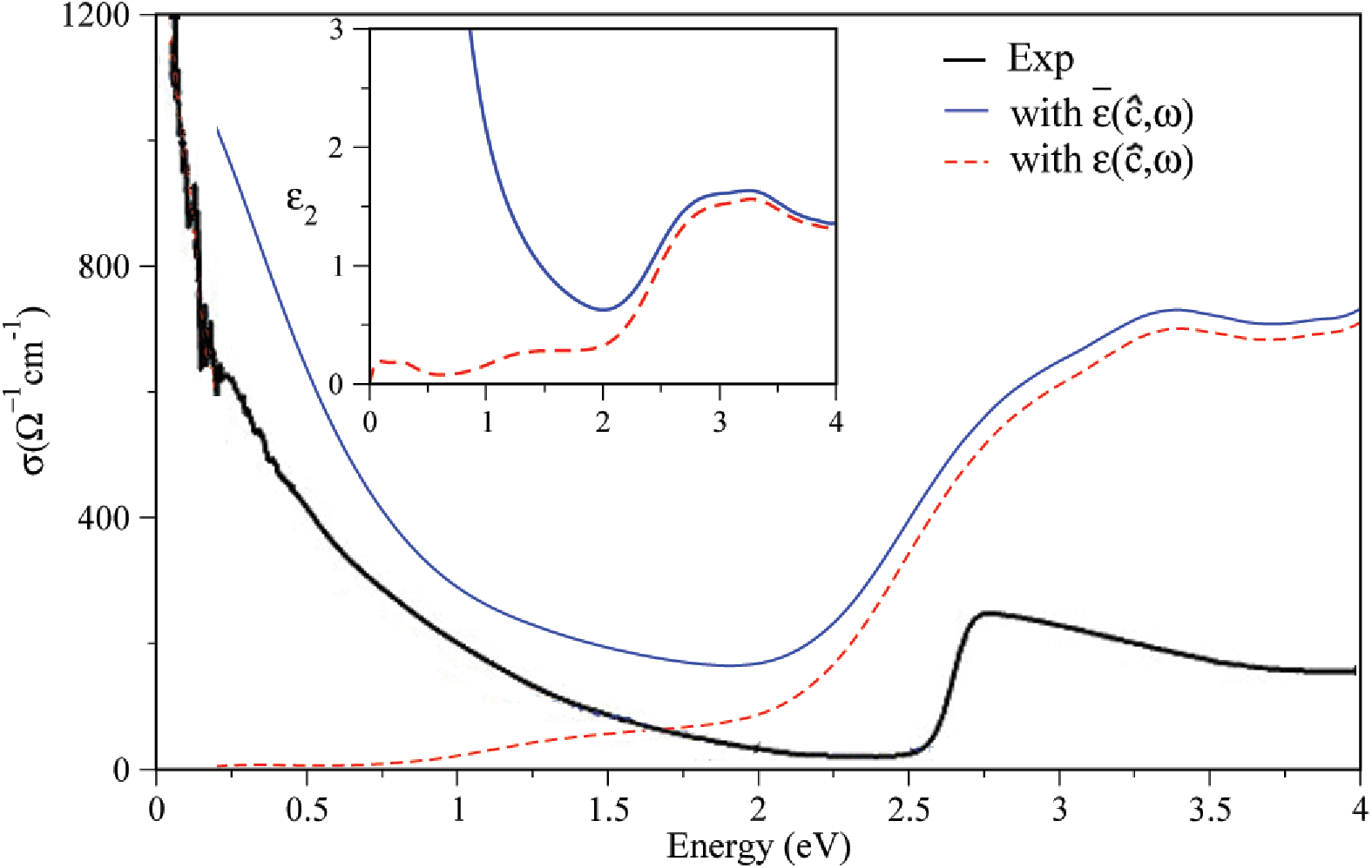}
\caption{\label{fig4} Measured spectrum of the optical
conductivity of PANI-CSA salts\cite{nature06} compared with $\sigma(\hat c;\omega)$
calculated with and without the Drude term. The inset shows the
$\omega$-dependence of $\epsilon_2(\hat c;\omega)$ and
${\overline\epsilon_2}(\hat c;\omega)$ in the region of
interest.}
\end{figure}

In Fig.~\ref{fig4}, we compare the measured spectrum of the optical conductivity\cite{nature06} with the spectra calculated with and without the Drude term.
In this case, the \lq\lq theoretical" curve overestimates appreciably the
experimental one and gives only a qualitative picture of the
spectral profile. To improve the quality of the calculated spectrum, one should probably include a larger number of $k-$points in the calculation of
the dielectric function, a request that is beyond our actual
computational capabilities.

%
%      Figura 4
%

\begin{figure}[ht]
\centering
\includegraphics[width=0.8\linewidth]{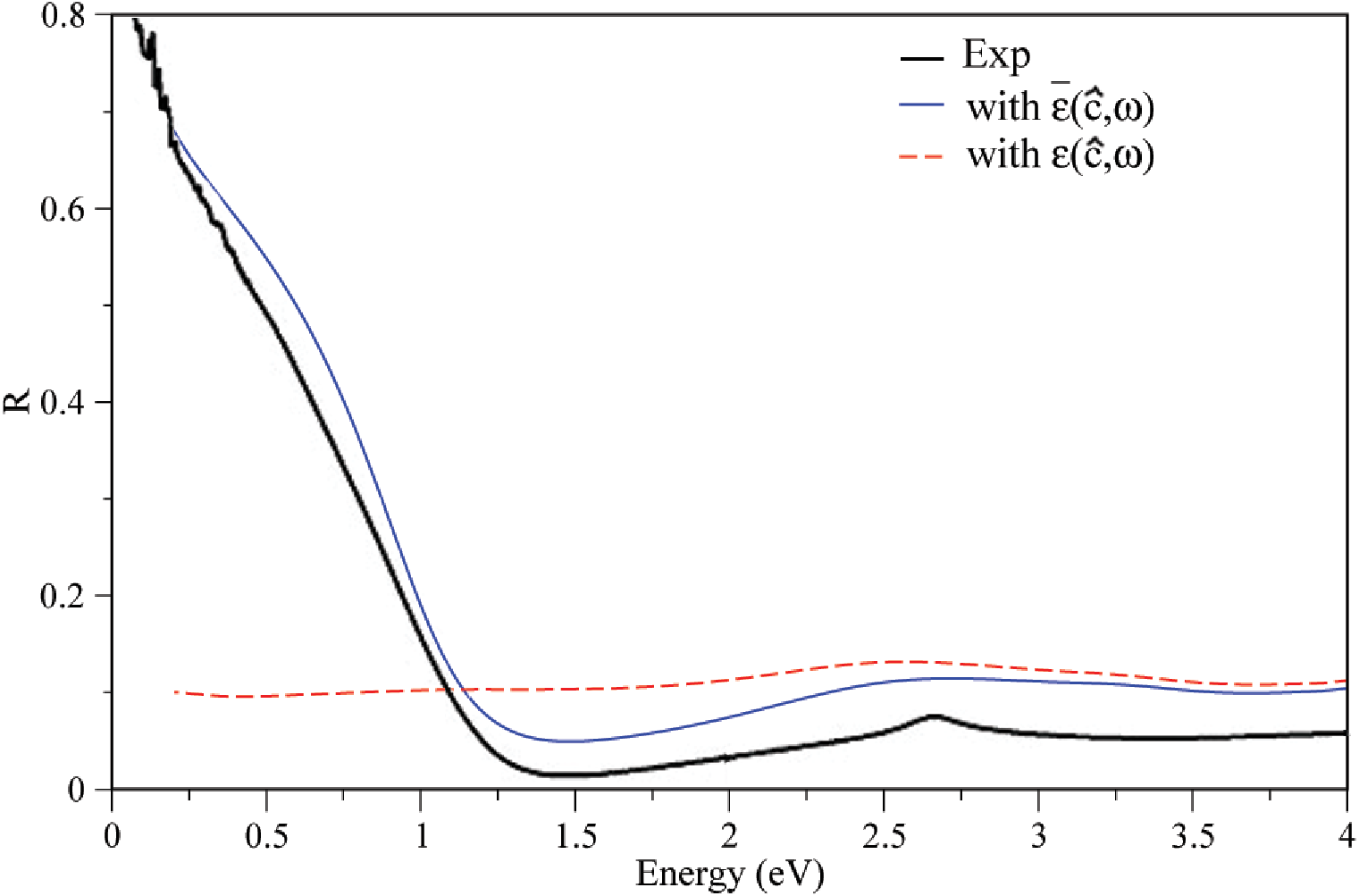}
\caption{\label{fig5} Measured spectrum of the reflectance of
PANI-CSA salts\cite{nature06} compared with $R(\hat c;\omega)$
calculated with and without the Drude term.}
\end{figure}

Finally, in Fig.~\ref{fig5}, we compare the measured spectrum of
reflectance \cite{nature06} with the spectra calculated with and
without the Drude term. We see,
again, a remarkable agreement between theory and experiment: the
\lq\lq theoretical" curve reproduces the characteristic minimum
around 1.4 eV, ascribed to the free-carrier plasma resonance
typical of a metal \cite{nature06}, and the bump in the region around 2.7 eV
due to inter-band transitions.

%
%      Figura 5
%

\section{Conclusions}\label{conc}
We have shown the possibility of calculating optical spectra of good quality for
crystalline polymers obtained by chemical protonation of conjugated polymers. We have used KS wavefunctions and energies implemented in the quantum-mechanical
expression of the macroscopic transverse dielectric function in the framework of the band
theory without the electron-hole interaction. To take into account also contributions of the intra-band transitions, we have added, to the dielectric function calculated ab initio,
a Drude-like term appropriate for metallic systems.

The calculated spectra are in remarkable agreement with the experiment and allow one to identify  contributions of inter- and the intra-band electronic transitions to the optical properties of these polymers. {\bf The analysis of the single-particle energy bands makes also evident the influence of the crystalline structure of the polymer: because of the specific crystal symmetry with two chains per unit cell, all bands are doubled and inter-chain transitions give a relevant contribution to the optical spectra. This fact confirms the need of a fully 3D approach to the study of these polymeric systems.}

We have also shown that absorption spectrum and other important optical
properties of PANI salts obtained from chemical doping of
conjugated polymers are not directly influenced by the type of
counterions produced in the acidification process. The
crystalline structure of ES-II with Cl$^-$ counterions in polaronic positions
represents indeed a structure appropriate also for
calculating optical properties of other PANI salts.

\section{Acknowledgments}
{\bf We thank Giuseppe Grosso for helpful discussions and suggestions.}
This work has been supported by MURST-PRIN 2004. We acknowledge the
allocation of computer resources from CNR-INFM \lq\lq Parallel
Computing Initiatives" and Cineca Supercomputing Center.

\end{document}